# Damage of bilayer structure in La$_3$Ni$_2$O$_{7-\delta}$ induced by high pO$_2$ annealing

Yulin Zhang,[1] Cuiying Pei,[2] Ning Guo,[3] Longlong Fan,[4] Mingxin Zhang,[2] Lingzhen Wang,[1] Gongting Zhang,[1] Feiyu Li,[1] Yunong Wang,[1] Chao Ma,[1] Wenyong Cheng,[1] Shanpeng Wang,[1] Qiang Zheng,[3*] Yanpeng Qi,[2*] and Junjie Zhang[1*]

[1]Institute of Crystal Materials, State Key Laboratory of Crystal Materials, Shandong University, Jinan 250100, Shandong China
[2]School of Physical Science and Technology, ShanghaiTech Laboratory for Topological Physics, Shanghai Key Laboratory of High-resolution Electron Microscopy, ShanghaiTech University, Shanghai 201210, China
[3]CAS Key Laboratory of Standardization and Measurement for Nanotechnology, CAS Center for Excellence in Nanoscience, National Center for Nanoscience and Technology, Beijing 100190, China
[4]Institute of High Energy Physics, Chinese Academy of Sciences, Beijing 100049, China

*Email: zhengq@nanoctr.cn (Q.Z.), qiyp@shanghaitech.edu.cn (Y.Q.) and junjie@sdu.edu.cn (J.Z.)

**Abstract**: The discovery of superconductivity with onset temperature of ~80 K in pressurized bilayer Ruddlesden-Popper La$_3$Ni$_2$O$_{7-\delta}$ has attracted much attention. Despite intense research, determination of the exact oxygen content and understanding of the relationship between superconductivity and oxygen content remain a big challenge. Here, we report a systematical study on the structure and physical properties of La$_3$Ni$_2$O$_{7-\delta}$ polycrystalline powders which were prepared using sol-gel method at ambient pressure and then annealed under various oxygen pressure. We found that high pO$_2$ annealing with slow cooling results in a new phase, which can be modeled using the hybrid single-layer-trilayer La$_3$Ni$_2$O$_7$ or the tetragonal bilayer La$_3$Ni$_2$O$_7$. Scanning transmission electron microscopy (STEM) measurements revealed significant single layers and trilayers after high oxygen pressure annealing, evidencing damage of the bilayer structure. The superconducting transition under high pressure became weak for high pO$_2$ annealed samples, which is consistent with the damage of the bilayer structure. Our results reveal that the bilayer structure is fragile and post-annealing under near atmosphere pressure of oxygen is suitable to maintain bilayer structure and increase oxygen content at the same time.

# 1. Introduction

The mechanism of high temperature superconductivity remains a big challenge in the fields of condensed matter physics and materials science[1-10]. In the electronic phase diagram of cuprates, besides superconductivity, the existence of competing phases including antiferromagnetic insulator, pseudogap, strange metal and Fermi liquid, and intertwined order such as charge density wave and spin density wave complicates the problem[4,6,11-18]. Design and synthesis of new high temperature superconductors in non-copper transition metal oxides may provide new materials platforms and new insights into unlocking the mystery of high temperature superconductivity.

Nickel is a neighbor of copper in the periodic table. Nickelates have been long sought for superconductivity[19-22]. Perovskite nickelates $RNiO_3$ (R=Pr-Lu, Y) and single-layer Ruddlesden-Popper (R-P) nickelates are model systems for studying metal-insulator transition and charge/spin stripes, respectively[23-27]. At ambient pressure, the bilayer and trilayer R-P nickelates are metals, showing a metal-to-metal transition upon cooling[28-30]. By analog, $Ni^{1+}$ with $d^9$ is isostructural to $Cu^{2+}$, which is the active ion in the high temperature superconducting cuprates. Theoretically, Anisimov et al.[31] predicted that square planar coordinated $Ni^{1+}$ can be doped to become a superconductor. In contrast, Lee et al.[32] found that $Ni^{1+}$ is not $Cu^{2+}$ based on theoretical calculations. Experimentally, Zhang et al. synthesized single crystals of square planar trilayer nickelates and found similarities between nickelates and cuprates including large orbital polarization[33], which is believed to be key ingredients to high temperature superconductivity. Superconductivity was finally discovered by Li et al in Sr-doped $NdNiO_2$ thin films, initiating the Nickel Age of superconductivity[21]. Later, high-$T_c$ superconductivity (onset ~ 80 K) was discovered by Sun et al. in pressurized bilayer $La_3Ni_2O_7$,[34] stimulating intense interest in Ruddlesden-Popper (RP) nickelates in both experiments[35-49] and theory[50-62].

Despite great efforts devoted to these materials, certain fundamental issues remain to be addressed. These include: (i) *Synthesis of high-quality single crystals*. Sample issues[30,34,36,41-43,45,48-49] including inhomogeneity, intergrowth and polymorphism have been reported in bilayer nickelate single crystals grown by the floating zone technique, hindering the understanding of the underlying physics surround nickelate superconductors. Recently, successful growth of the bilayer nickelate single crystals at ambient pressure using flux method provide a route to overcome this issue[43,63-64]; (ii) *The relationship between superconductivity and oxygen content*. Oxygen content plays an important role in tuning electrical properties of $La_3Ni_2O_{7-\delta}$[48,53,65-66]. Theoretical studies have shown that the inner apical oxygen $2p_z$ orbitals bridging the two neighboring $NiO_2$ planes are involved in the interlayer super-exchange interaction and thus affect the superconductivity[52-60]. Due to the existence of impurities and intergrowth of R-P phases, determination of exact oxygen content in $La_3Ni_2O_{7-\delta}$ is very challenging[61-62]; (iii) *Bulk or filamentary superconductivity in nature*. Zhou et al.[67] reported low superconducting volume fraction and Wang et al.[47] reported the absence of diamagnetic signal in $La_3Ni_2O_7$, consistent with filamentary superconductivity. In sharp contrast, Li et al.[68] reported maximum superconducting volume fraction of 48% at 19.4 GPa in $La_3Ni_2O_7$, implying bulk superconductivity. Similar debate also exist in the trilayer nickelate $Pr_4Ni_3O_{10}$[38,69]. Recent report[44] of zero resistance starting at 60 K and superconducting volume fraction of >90% in Pr-substituted $La_2PrNi_2O_7$ provides a direction to achieve bulk superconductivity; (iv) *Is it possible to achieve ambient pressure nickelate superconductivity in bulk samples – polycrystalline powders and single crystals?* Recently superconductivity up to 48 K has been found in ultrathin films at ambient pressure by mimicking crystal structure in superconducting state under high pressure[70]. By substituting La with smaller rare earth elements, Li et al.[63] successfully grew a series of nickelate single crystals at ambient pressure with smaller in-plane lattice constants and achieved superconductivity above 90 K under high pressure. Unfortunately, substitution using smaller rare earth elements is still not enough to drive superconductivity to ambient pressure.

In this contribution, we carried out systematic research on post-annealing $La_3Ni_2O_{7-\delta}$ polycrystalline powders under various oxygen pressure in order to tune the electronic structure of $La_3Ni_2O_{7-\delta}$, with the hope of obtaining an electronic phase diagram of $T_c$ vs oxygen content. Unexpectedly, by combining of powder X-ray diffraction (PXRD), scanning transmission electron microscopy (STEM) and pair distribution function (PDF) measurements, we found that high $pO_2$ annealed samples were subjected to structural damage, which is consistent with the weak superconducting transition under high pressure. Our results suggest that oxygen engineering requires near ambient pressure to preserve crystalline integrity.

## 2. Results and discussion
## 2.1 Structural response to various $pO_2$ annealing
### 2.1.1 Average structure via PXRD

Polycrystalline powders of $La_3Ni_2O_{7-\delta}$ were prepared in air using the sol-gel method. **Figure 1a** shows the powder X-ray diffraction pattern of as-synthesized sample. No impurity peaks are found. Rietveld refinement was performed using *Amam* structure as a starting model ($La_3Ni_2O_7$-2222), which converged to $R_{exp}$ = 2.89%, $R_{wp}$ = 8.46%, $R_p$ = 5.77%, and GOF = 2.93 with cell parameters of $a$ = 5.39885(16) Å, $b$ = 5.45130(16) Å, and $c$ = 20.5163(6) Å. The intensities around 2θ = 26° and 44°, which correspond to (006) and (00<u>10</u>), respectively, are not well fitted by the *Amam* structural model. This indicates defects including stacking faults in the as-synthesized samples, similar to the report by Wang [44].

We then annealed the samples in order to increase oxygen content. The annealing conditions include oxygen pressure ($pO_2$=1, 19, 37, 77, and 100 bar), dwelling temperature (500 °C), dwelling time (3 days), and cooling time of 1 day vs 30 min from 500 to 30 °C. **Figure 1b** shows the PXRD pattern of the sample annealed at 1 bar $O_2$. The Rietveld refinement converged to $R_{exp}$ = 3.40%, $R_{wp}$ = 10.50%, $R_p$ = 7.41%, and GOF = 2.84 with cell parameters of $a$ = 5.4016(3) Å, $b$ = 5.4542(3) Å, and $c$ = 20.5222(13) Å. As can be seen, the diffraction pattern is quite similar to the as-synthesized one. **Figure 1c** shows the pattern of the sample annealed under 77 bar $O_2$ with cooling time of 1 day. Unexpected, extra reflection peaks appear, in particular at 2θ ~ 45° and 47.5°. These extra peaks do not match those in the database (PDF4+-2025). To understand this new phase, we then carried out Le Bail fit using the $La_3Ni_2O_7$-2222 model together with different $La_3Ni_2O_7$-1313 models, the $La_4Ni_3O_{10}$ model and other possible RP phases (**Figure S1-S4**). Among them, the best fit is from the combination of $La_3Ni_2O_7$-2222 (*Amam*) and $La_3Ni_2O_7$-1313 (*Cmmm*), as shown in **Figure S1**. We then performed Rietveld refinement using this combination. The refinement converged to $R_{exp}$ = 3.67%, $R_{wp}$ = 10.57%, $R_p$ = 7.40%, and GOF = 2.88 with cell parameters of $a$ = 5.4544(10) Å, $b$ = 5.4620(10) Å, and $c$ = 20.227(2) Å for $La_3Ni_2O_7$-1313 (*Cmmm*) and $a$ = 5.4057(6) Å, $b$ = 5.4572(7) Å, and $c$ = 20.496(2) Å for $La_3Ni_2O_7$-2222 (*Amam*). The obtained lattice parameters are indeed consistent with that of $La_3Ni_2O_7$-1313 (*Cmmm*), suggesting that a phase transition from bilayer phase to the hybrid single-layer and trilayer polymorph. The obtained mass ratio is 49.9 wt.% for $La_3Ni_2O_7$-2222 (**Figure 1c**). In addition to Rietveld refinement, we fitted the characteristic peak at ~47.5° using two pairs of peaks ($K_{\alpha 1}$ and $K_{\alpha 2}$ for phase 1 and 2). In this way, we can estimate the mass ratio of the two phases (**Figure S5b**), and 42.3 wt.% was obtained for $La_3Ni_2O_7$-2222, which is close to the value from Rietveld refinement. We also carried out annealing with different cooling procedure, we found that the quantity of the new phase decreases with decreasing cooling time. This result indicates that the appearance of the new phase occurs during cooling. However, we haven't succeeded in obtaining single phase of this new phase. Recently, Shi et al.[64] reported a tetragonal phase of $La_3Ni_2O_7$ with similar $c$ axis. Considering that it is also obtained via annealing at high $pO_2$, we then refined our diffraction pattern using this tetragonal model (**Figure 1d**). The Rietveld refinement converged to $R_{exp}$ = 3.67%, $R_{wp}$ = 8.69%, $R_p$ = 6.31%, and GOF = 2.36 with cell parameters of $a$ = 3.8574(2) Å, and $c$ = 20.2289(15) Å for $La_3Ni_2O_7$ (*I4/mmm*) and $a$ = 5.4005(5) Å, $b$ = 5.4535(5) Å, and $c$ = 20.5162(19) Å for $La_3Ni_2O_7$-2222 (*Amam*). The obtained lattice parameters are pretty close to the report by Shi et al.[64] The $R_{wp}$ is slightly smaller than that of $La_3Ni_2O_7$-1313 (*Cmmm*). Both models can explain our experiment, and further experiments are needed to distinguish them.

To avoid impurities after annealing, we optimized our annealing conditions. For these experiments, we annealed the samples at 500 °C for 3 days and then released oxygen pressure quickly, followed by quenching the sample (cooling time was 10 min from 500 to 300 °C). **Figure 1e** presents the diffraction pattern of such a sample. No impurity peaks or peaks of the new phase are seen. The Rietveld refinement using $La_3Ni_2O_7$-2222 (*Amam*) converged to $R_{exp}$ = 3.95%, $R_{wp}$ = 10.46%, $R_p$ = 7.44%, and GOF = 2.65 with cell parameters of $a$ = 5.4051(4) Å, $b$ = 5.4537(4) Å, and $c$ = 20.5214(16) Å. **Table I** summarizes the result of $La_3Ni_2O_7$ samples annealed at various $pO_2$ and cooling procedure. Among them, the as-synthesized samples and samples annealed at high oxygen pressure followed by pressure releasing quickly and quenching are single phase of $La_3Ni_2O_7$-2222, while the samples annealed at high $pO_2$ followed by slowing cooling consist of $La_3Ni_2O_7$-2222 and a new phase. Whether this new phase is hybrid $La_3Ni_2O_7$-1313 (*Cmmm*) or tetragonal bilayer $La_3Ni_2O_7$ requires further characterization.

### 2.1.2 Local Structure via STEM

In order to understand the new phase we obtained during annealing, we performed real-space imaging of as-synthesized sample (S1 in **Table 1**), annealed sample S3 and S4 via STEM. For as-synthesized polycrystalline powders, stacking of bilayers is observed as expected, with tiny intergrowth of R-P phases (**Figure S6**). This is consistent with our PXRD result and the previous reports[44]. In sharp contrast, the sample annealed at 77 bar with cooling time of 1 day show a large amount of short-range intergrowth of R-P phases instead of long-range ordered hybrid $La_3Ni_2O_7$-1313 (**Figure 2 and Figure S7**), indicating that the bilayers are damaged during annealing. The cracking of bilayers to form trilayers and single-layers resulted in a much shorter $c$ axis, in good agreement with our PXRD analysis. We also performed STEM on the sample annealed at 77 bar with pressure releasing quickly and quenching to room temperature from 500 °C, as shown in **Figure S8**. As can be seen, most areas are dominated by bilayers, with a small amount of single-layers and trilayers. Overall, short-range intergrowth is worse than the as-synthesized sample but much better than the sample annealed at high pressure with cooling time of 1 day. Our result indicates that high $pO_2$ annealing of the $La_3Ni_2O_{7-\delta}$ samples compromises the integrity of the bilayer structure to form single layers and trilayers.

### 2.1.3 Local structure via PDF

To further understand the local structure, we performed pair distribution function measurements on as-synthesized sample (S1), annealed sample at 19 bar followed by pressure releasing quickly and quenching (S5, see **Figure S9** for PXRD and Rietveld refinement), and annealed sample at 100 bar with cooling time of 1 day from 500 to 30 °C (S7, see **Figure S10** for PXRD and Rietveld refinement). **Figure 3a** shows the $G(r)$ for three samples in the range of 20-35 Å. Clearly, the $G(r)$ of S7, which contains significant new phase, is remarkably different from that of S1 and S5, indicating very different local environments between S7 and S1/S5. We first look at the data in the range of 3-5 Å. Interestingly, the $G(r)$ for three samples are essentially identical. The refinements using $La_3Ni_2O_7$-2222 converged to $R_{wp}$ = 2.09%, $R_{wp}$ = 1.89%, $R_{wp}$ = 2.35%, respectively. This indicates that the bond lengths and coordination numbers of the nearest-neighbor atomic pairs in the different samples are similar (**Figure 3b**). We then performed refinement for the range of 5-40 Å (**Figure 3c**). The refinements of S1 and S5 using $La_3Ni_2O_7$-2222 converged to $R_{wp}$ = 15.65% and $R_{wp}$ = 15.96%, respectively. The refinement of S7 using $La_3Ni_2O_7$-2222 and $La_3Ni_2O_7$-1313 converged to $R_{wp}$ = 14.1%. In addition, the refinement of S7 using $La_3Ni_2O_7$-2222 and $La_3Ni_2O_7$ ($I4/mmm$) converged to $R_{wp}$ = 14.1%, see **Figure S11**. In contrast, the refinements of S7 using $La_3Ni_2O_7$-1313 converged to $R_{wp}$ = 23.28%, and using $La_3Ni_2O_7$ ($I4/mmm$) converged to $R_{wp}$ = 20.79%, see **Figure S12-13**. Our refinement cannot differentiate $La_3Ni_2O_7$-2222+$La_3Ni_2O_7$-1313 and $La_3Ni_2O_7$-2222+$La_3Ni_2O_7$ ($I4/mmm$). However, a significant deviation from those of S1 is observed, i.e., the $c$-axis was refined to 20.1896 Å in $La_3Ni_2O_7$-1313 and 20.4055 Å in $La_3Ni_2O_7$-2222, which is much smaller than 20.5173 Å in S1. The obtained $c$ axis is well consistent with our Rietveld refinement on PXRD data.

## 2.2 Ambient-pressure physical properties
### 2.2.1 Electrical resistivity of samples annealed under various $pO_2$

We carried out electrical transport measurements on the samples annealed at various $pO_2$ and cooling procedure from two batches of $La_3Ni_2O_7$ samples (**Figure 4**). Since high oxygen pressure annealing with 1 day cooling time resulted in a new phase with bilayer damaged, we focus on the resistivity of samples annealed at high oxygen pressure with pressure releasing quickly and quenching. The resistivity of the as-synthesized sample at ambient pressure increases first and then decreases with decreasing temperature, reaching a minimum at around 125 K. Such an anomaly has been proposed to correspond to a density wave transition[65,71-72]. Notably, the resistivity of 1 bar annealed sample is more metallic than the as-synthesized samples. With increasing annealing $pO_2$, overall, the resistivity became more metallic, with smaller upturn at low temperature. This improvement may be attributed to the reduction of oxygen vacancies. We attempted to determine precise oxygen content using TGA but obtained not-so-meaningful result due to the existence of intergrowth, please see **Figure S14-16 and Table S1** for more details.

### 2.2.2 Magnetic susceptibility of samples annealed under various $pO_2$

The temperature dependent magnetic susceptibility data of $La_3Ni_2O_{7-\delta}$ before and after oxygen pressure annealing in the temperature range of 1.8-300 K are shown in **Figure S17** in the supplementary information. The sample was rapidly cooled down to 1.8 K under zero field and a field of 0.4 T was applied. Data were collected on warming. The magnetic susceptibility decreases with decreasing temperature in the range of 300-70 K, showing the characteristic of the low dimensional electron systems[72]. Below 70 K, paramagnetic behavior is observed. Our result is consistent with the previous reports [48,71]. Annealing at various $pO_2$ basically does not change the magnetic properties.

## 2.3 High-pressure electrical resistivity of samples annealed under various $pO_2$

The temperature dependence of the resistance R(T) is displayed in **Figure 5** for various high pressures. Resistance under high pressures was measured using a nonmagnetic diamond anvil chamber (DAC) and the van der Pauw method. Increasing pressure induces a continuous suppression of the overall magnitude of resistance. At higher pressure region, a pressure-induced insulator-metal transition was obtained accompanied by a drop at around 80 K at low temperature. Upon further increasing the pressure, the drop of R(T) becomes more pronounced, indicating the emergence of a superconducting transition. Zero resistance is not observed and the upward curvature of the resistivity at low temperature may be due to the local magnetic moment caused by the magnetic impurities contained in the sample. With the increase of pressure, the superconducting transition temperature gradually moves to the low temperature, which is consistent with the previous report [34]. It should be noted that the $T_c$ onset is almost unchanged compared with the samples annealed at different oxygen pressures. The superconducting transition became weak for high $pO_2$ annealed samples, which is ascribed to the damage of the bilayer structure in $La_3Ni_2O_{7-\delta}$ sample.

## 3. Conclusion

Single phase of $La_3Ni_2O_{7-\delta}$ polycrystalline powder was prepared by sol-gel method and then annealed under oxygen pressure using three ways ($pO_2$=1bar and slow cooling, high $pO_2$ with slow cooling, high $pO_2$ with pressure releasing quickly and quenching). We found that annealing using the first and third way resulted in single phase of $La_3Ni_2O_7$-2222. Unexpectedly, a new phase was obtained using the second annealing method. A combined analysis on the samples using the second annealing method using PXRD, STEM and PDF revealed that the bilayer structure was damaged, which is consistent with a weaker superconducting transition in the high-pressure resistance measurements. Our results reveal that near-atmospheric oxygen pressure annealing is suitable for increasing oxygen content while maintaining the bilayer structure, which is the basic structural unit for high-temperature superconductivity.

**Note added**: During the preparation of this manuscript, we noticed two independent manuscripts reporting the effect of post annealing. Shi et al. reported a tetragonal phase of $La_3Ni_2O_7$ obtained via annealing at high $pO_2$ and the absence of superconductivity at ambient pressure and high pressure[64]. The lattice parameters of this tetragonal phase are similar to what we obtained in our annealing experiment. In contrast, Huo et al. reported signature of 80 K superconductivity at ambient pressure at high $pO_2$ annealed $La_3Ni_2O_7$ with a superconducting volume fraction of ~0.2%[74]. We haven't obtained diamagnetic signal in our annealed samples.

**Acknowledgement**
Work at Shandong University was supported by the National Natural Science Foundation of China (12074219 and 12374457), the TaiShan Scholars Project of Shandong Province (tsqn201909031), and the QiLu Young Scholars Program of Shandong University. Work at ShanghaiTech University was supported by the National Natural Science Foundation of China (52272265) and the National Key R&D Program of China (2023YFA1607400). J.Z. thanks Dr. Xutang Tao from Shandong University, Dr. Yu-Sheng Chen and Dr. Tieyan Chang from The University of Chicago for stimulating discussions.

**Author contributions**
J.Z. conceived the project; Y.Z. synthesize powders, performed the powder X-ray diffraction experiments, carried out magnetic susceptibility and transport measurements at ambient pressure with the help of L.W, G.Z., F.L., Y.W., C.M., W.C., S.W. and J.Z.; C.P., M.L. and Y.Q. performed high pressure measurements and data analysis; N.G. and Q.Z. carried out STEM measurements; Y.Z. and J.Z. wrote the draft with contributions from all coauthors.

**Competing interests**

The authors declare no conflict of interest.

**Experimental section**

**Synthesis.** The Polycrystalline $La_3Ni_2O_{7-\delta}$ powder samples were prepared using the sol-gel method[47,65-66]. Stoichiometric amounts of NiO (Sigma Aldrich, 99.99%) and $La_2O_3$ (Sigma Aldrich, 99.99%) were added to a quartz crucible and dissolved in deionized water and nitric acid by heating and stirring until it becomes a clear green solution, citric acid was added and heated and stirred to form a green gel. The resulting gel was dried and sintered at 400°C for 5 hours to produce a fly ash-like black powder. This powder was then pressed into a sheet and sintered in an alumina crucible at 1100-1150 °C for 12 hours until a pure phase $La_3Ni_2O_7$ polycrystalline powder was obtained.

**Oxygen pressure annealing.** The $La_3Ni_2O_{7-\delta}$ polycrystalline powder was annealed at 500 °C in an alumina crucible using a Xi'an Taikang furnace at different oxygen pressures. During annealing under high oxygen pressure, it was found that different cooling time and pressure releasing methods have different result, see details in Table I. We found that using the cooling time of 10 min from 500 to 300 °C with pressure release quickly and quenching, the annealed $La_3Ni_2O_{7-\delta}$ remain 2222. Otherwise, a new phase with bilayer damaged appear.

**Characterization.** The powder X-ray diffraction (PXRD) data of samples were collected using a Bruker AXS D2 Phaser diffractometer at room temperature and ambient pressure with Cu $K_\alpha$ radiation ($\lambda$ = 1.5418 Å) in the 2θ range of 10º-90º. The scan step size was 0.02 º with a scan time of 0.1 s per step. The atomic pair distribution function (PDF) data of samples was collected from the Institute of High Energy Physics, Chinese Academy of Sciences (IHEP, CAS). High-angle annular dark field-scanning transmission electron microscopy (HAADF-STEM) images were acquired at an accelerating voltage of 300 kV on a double-aberration corrected transmission electron microscope (Spectra 300, Thermo Fisher Scientific), equipped with a field-emission electron source in Chinese National Center for Nanoscience and Technology. The oxygen content of the sample was determined by calculating the mass difference before and after reduction in a 4% $H_2$/Ar using a Mettler-Toledo TGA/DSC$^{3+}$. The sample was measured in an alumina crucible and started at 100 °C, held for 1 hour, then heated to 900 °C, held for 2 hours, and finally cooled to 100 °C, held for 1 hour. Three blanks had been run previously to establish a good baseline. The final oxygen content was calculated using the mass at 900 °C. The resolution of our equipment is around 10 μg. The electrical resistance of the samples was performed at atmospheric pressure using the four-probe method in the temperature range of 1.8-300 K. Data were collected using a Quantum Design Physical Property Measurement System (PPMS). The magnetic susceptibility of samples was measured by a Quantum Design Magnetic Properties Measurement System (MPMS) under the magnetic field of 0.4 T in the temperature range of 1.8-300 K.

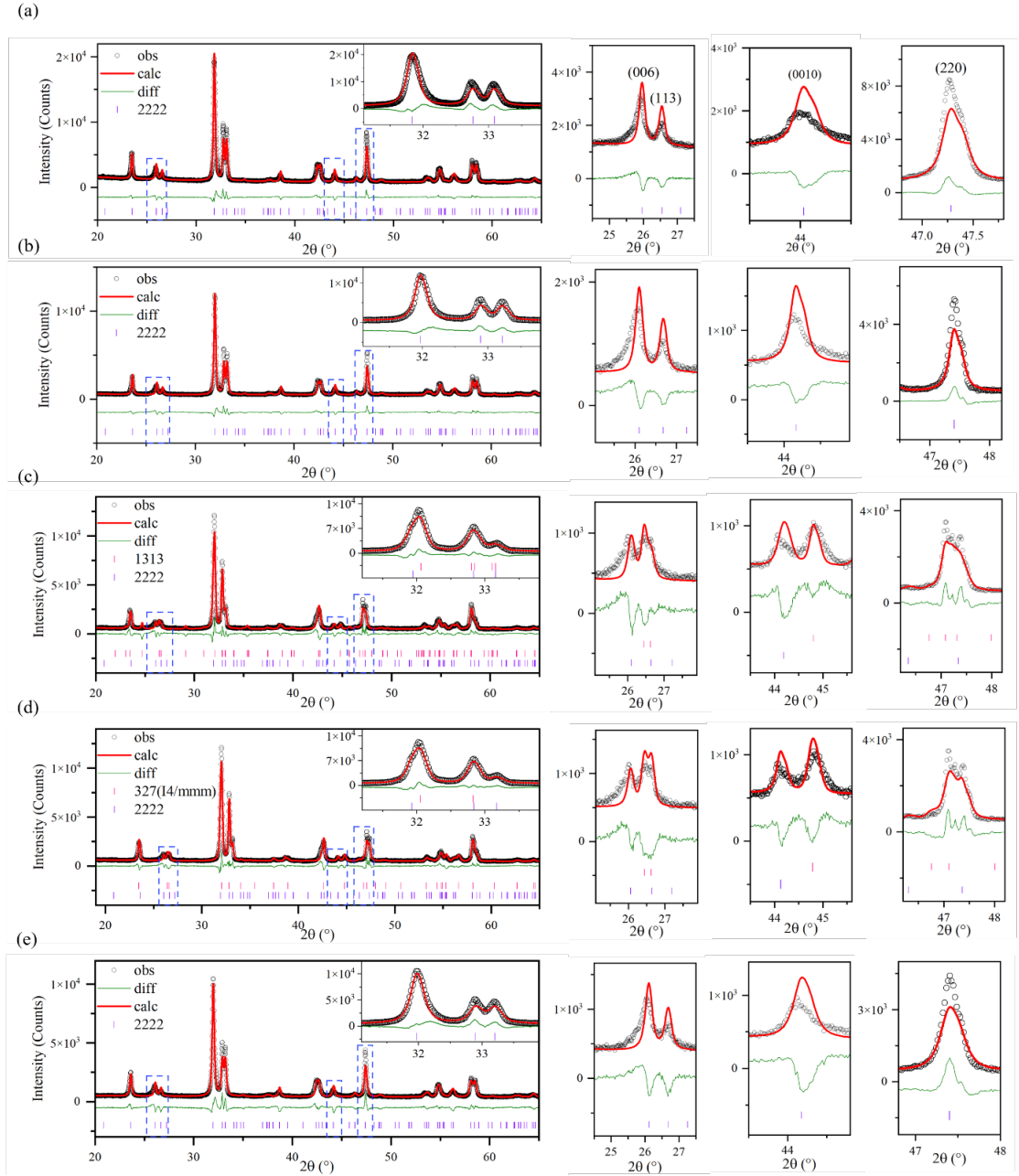

**Figure 1.** Powder X-ray diffraction patterns of La$_3$Ni$_2$O$_{7-\delta}$ samples and Rietveld refinement. (a) As-synthesized polycrystalline sample with refinement converged to $R_{exp}$ = 2.89%, $R_{wp}$ = 8.46%, $R_p$ = 5.77%, and GOF = 2.93. The inset shows the zoom in the range of 31°-34°. The right three figures show the zoom around 26°, 44° and 47.5°. (b) Annealed at 1 bar oxygen pressure for 3 days and the cooling time of 1 day from 500 to 30 °C. The refinement converged to $R_{exp}$ = 3.40%, $R_{wp}$ = 10.50%, $R_p$ = 7.41%, and GOF = 2.84. (c) Annealed at 77 bar oxygen pressure for 3 days with 1 day cooling time and Rietveld refinement using 1313 (*Cmmm*) and La$_3$Ni$_2$O$_7$-2222, which converged to $R_{exp}$ = 3.67%, $R_{wp}$ = 10.57%, $R_p$ = 7.40%, and GOF = 2.88. (d) Annealed at 77 bar oxygen pressure for 3 days with 1 day cooling time and Rietveld refinement using La$_3$Ni$_2$O$_7$ (*I4/mmm*) and La$_3$Ni$_2$O$_7$-2222, which converged to $R_{exp}$ = 3.67%, $R_{wp}$ = 8.69%, $R_p$ = 6.31%, and GOF = 2.36. The cooling time is 1 day from 500 to 30 °C. Additional reflection peaks that do not belong to bilayer La$_3$Ni$_2$O$_7$ are clearly seen. (e) Annealed at 77 bar oxygen pressure for 3 days with 10 min cooling from 500 to 300 °C with pressure releasing quickly and quenching, which converged to $R_{exp}$ = 3.95%, $R_{wp}$ = 10.46%, $R_p$ = 7.44%, and GOF = 2.65.

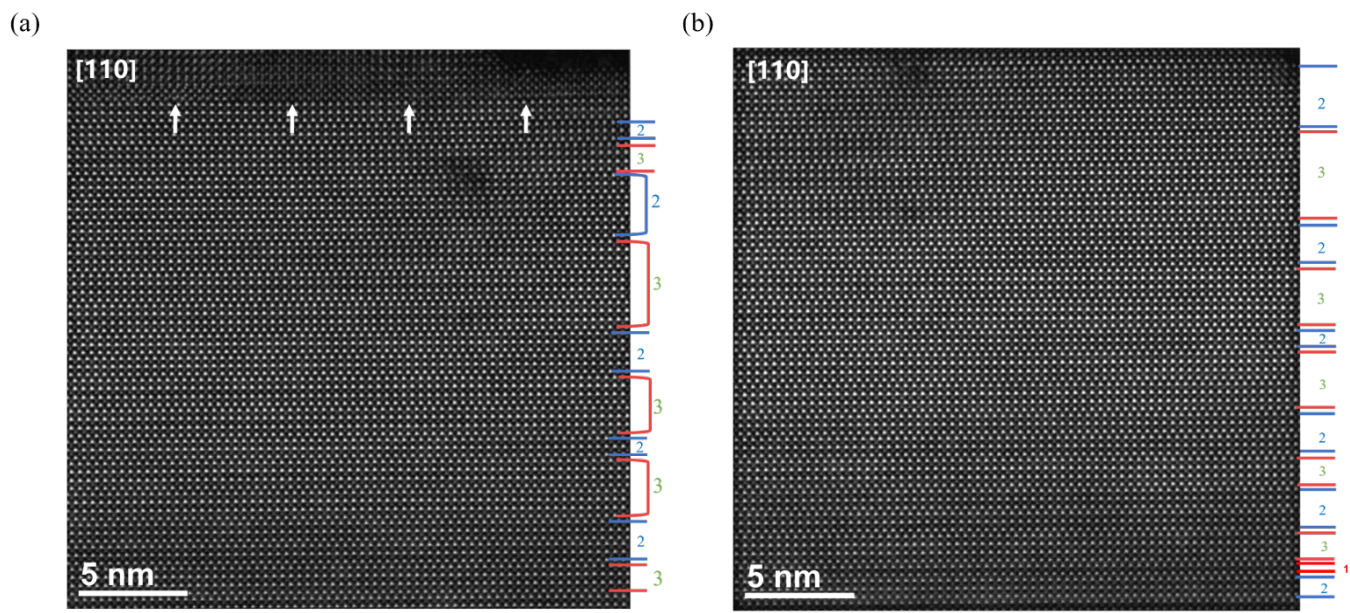

**Figure 2.** Typical atomic-scale HAADF-STEM images along the [110] orientation for samples annealed at 77 bar oxygen pressure for 3 days with slowing cooling (1 day from 500 to 30 °C). The numbers 1, 2, and 3 on the right side indicate single-layer $La_2NiO_4$, bilayer $La_3Ni_2O_7$ and trilayer $La_4Ni_3O_{10}$, respectively. (a) and (b) represent different regions. Above the white arrows, there are mixed rows of bilayer and trilayer blocks of $NiO_6$ octahedra.

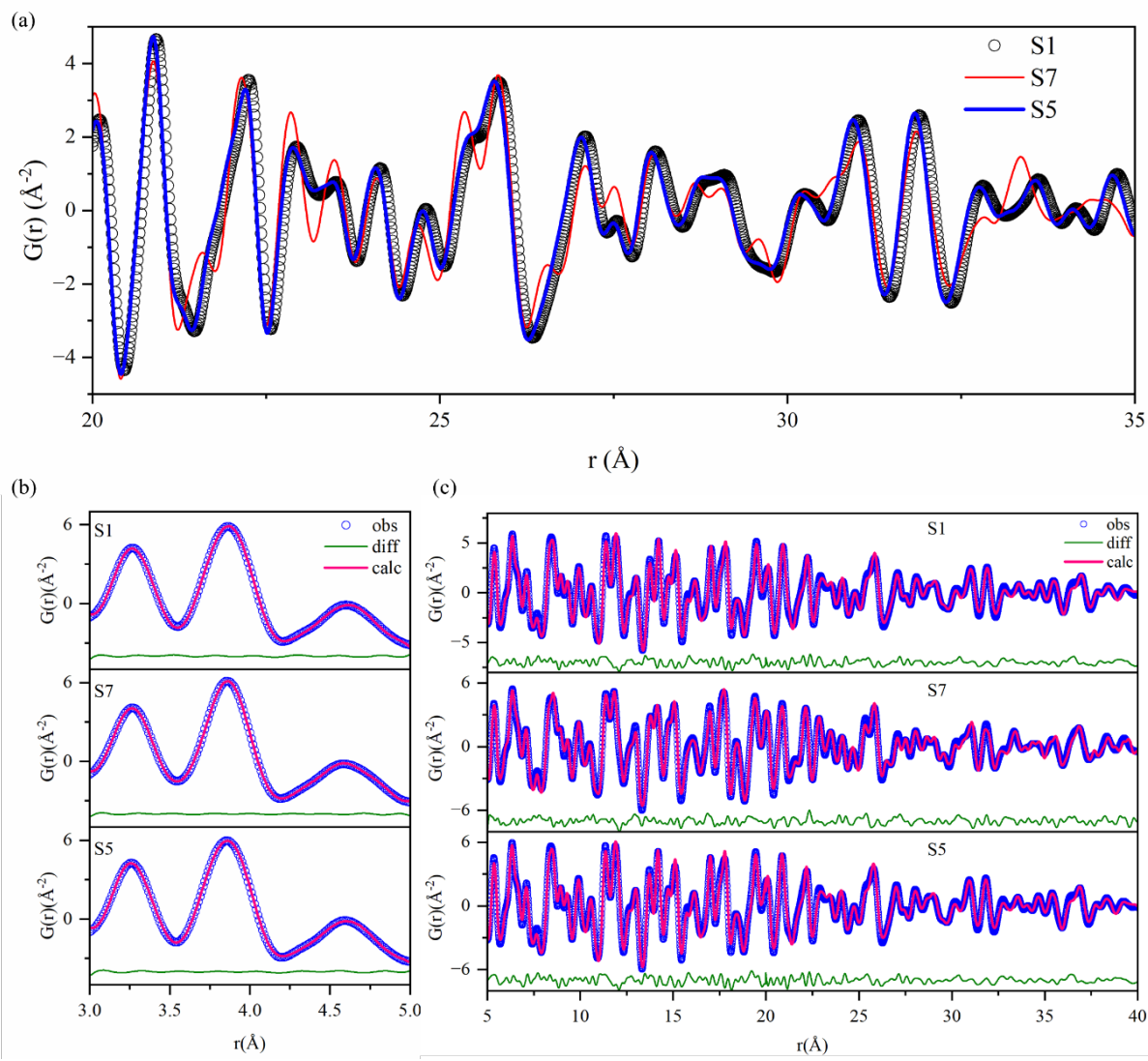

**Figure 3.** Pair distribution function data of as-synthesized $La_3Ni_2O_{7-\delta}$ samples (S1), annealed sample (S7), and annealed sample (S5). (a) Experimental data for three samples in the range of 20-35 Å. (b) and (c) Refinement results of PDF data for different ranges.

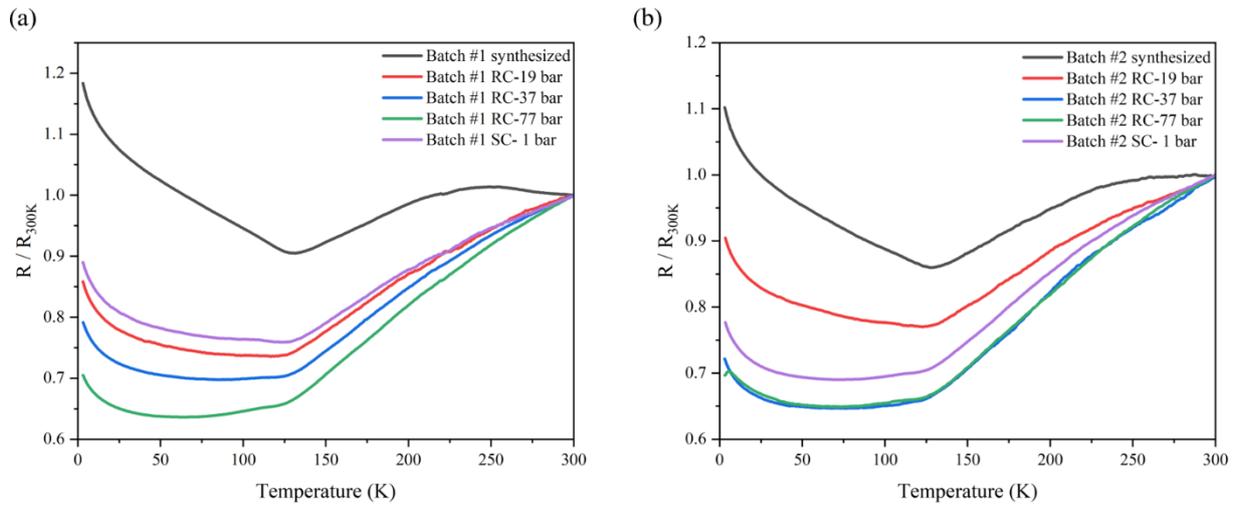

**Figure 4.** Temperature dependent resistance at atmospheric pressure for La$_3$Ni$_2$O$_{7-\delta}$ samples after annealing at different oxygen pressure. Note RC denotes rapid cooling (10 min from 500 to 300 °C with pressure releasing quickly and quenching, followed by furnace cooling to room temperature) and SC denotes slow cooling (1 day from 500 to 30 °C).

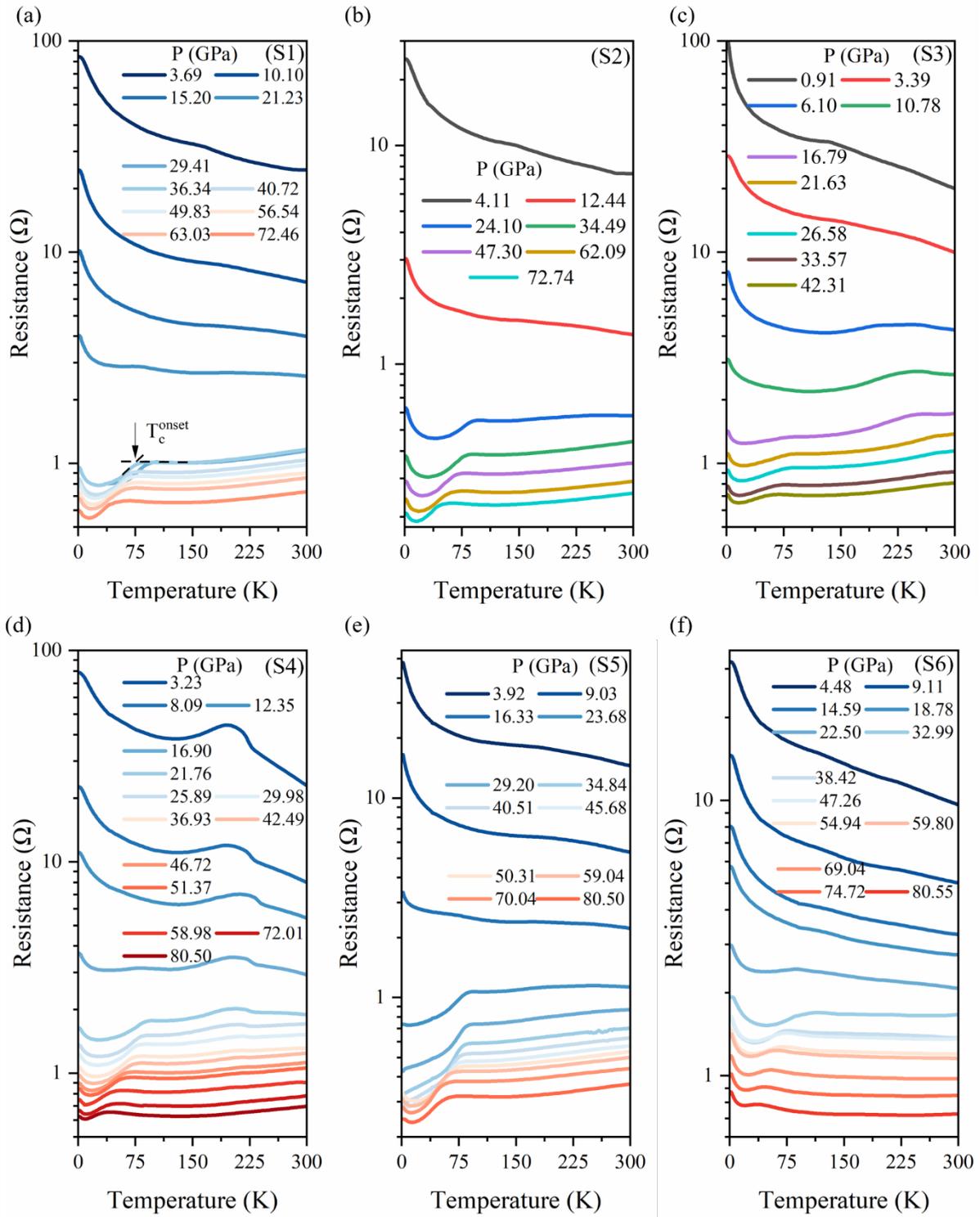

**Figure 5.** Pressure and temperature dependence of the electrical resistance R(T) of the La$_3$Ni$_2$O$_7$ polycrystalline samples annealed at different conditions. The T$_c^{onset}$ is determined as the interception between two straight lines below and above the superconducting transitions. (a), (b), (c), (d), (e), (f) are the data corresponding to S1, S2, S3, S4, S5, S6 in the sample information table (**Table I**), respectively.

**Table I.** Summary of $La_3Ni_2O_7$ samples with various annealing $pO_2$.

| No. | Synthesis conditions | Annealing conditions | PXRD | STEM | PDF | Transport properties |
|---|---|---|---|---|---|---|
| S1 | Sol-gel Sintered at 1150 °C for 12 h. | - | $La_3Ni_2O_7$-2222 | Mainly bilayer, minor intergrowth | $La_3Ni_2O_7$-2222 | Weakly insulating below 135 K at ambient pressure, and superconduct under high pressure. |
| S2 | Sol-gel Sintered at 1150 °C for 12 h. | $pO_2$ = 1 bar, T = 500 °C for 3 d, Cooling time is 1 d from 500 to 30 °C. | $La_3Ni_2O_7$-2222 | - | - | Weakly insulating below 135K at ambient pressure is suppressed, and superconduct under high pressure. |
| S3 | Sol-gel Sintered at 1150 °C for 12 h. | $pO_2$ = 77 bar, T = 500 °C for 3 d, 1 d from 500 to 30 °C. | $La_3Ni_2O_7$2222 + $La_3Ni_2O_7$-1313 or $La_3Ni_2O_7$-2222 + $La_3Ni_2O_7$-I4/mmm | Severe intergrowth | - | Metallic at ambient pressure, and superconduct under high pressure. |
| S4 | Sol-gel Sintered at 1150 °C for 12 h. | $pO_2$ = 77 bar, T = 500 °C for 3 d, pressure releasing quickly, 10 min from 500 to 300 °C, followed by furnace cooling. | $La_3Ni_2O_7$-2222 | Mainly bilayer, minor intergrowth | - | Metallic at ambient pressure except low-temperature upturn, and superconduct under high pressure. |
| S5 | Sol-gel Sintered at 1150 °C for 48 h. | $pO_2$ = 19 bar, T = 500 °C for 3 d, pressure releasing quickly, 10 min from 500 to 300 °C, followed by furnace cooling. | $La_3Ni_2O_7$-2222 | - | $La_3Ni_2O_7$-2222 | Metallic at ambient pressure except low-temperature upturn, and superconduct under high pressure. |
| S6 | Sol-gel Sintered at 1150 °C for 48 h. | $pO_2$ = 37 bar, T = 500 °C for 3 d, pressure releasing quickly, 10 min from 500 to 300 °C, followed by furnace cooling. | $La_3Ni_2O_7$-2222 | - | - | Weakly insulating is suppressed, and superconduct under high pressure. |
| S7 | Sol-gel Sintered at 1150 °C for 12 h. | $pO_2$ = 100 bar, T = 500 °C for 12 h, 1 d from 500 to 30 °C. | $La_3Ni_2O_7$-2222 + $La_3Ni_2O_7$-1313 or $La_3Ni_2O_7$-2222 + $La_3Ni_2O_7$-I4/mmm | - | $La_3Ni_2O_7$-2222 + $La_3Ni_2O_7$-1313 or $La_3Ni_2O_7$-2222 + $La_3Ni_2O_7$-I4/mmm | - |